\documentclass[aps,prl,reprint,twocolumn,showpacs,showkeys,superscriptaddress,bibliography]{revtex4-1}
\usepackage{amsmath,amssymb,amstext}
\usepackage[usenames,dvipsnames]{color}
\usepackage{graphicx}
\usepackage[english]{babel}
\usepackage{wasysym}
\usepackage{bm,bbold,braket}
\usepackage{natbib}
\usepackage{txfonts, comment, stmaryrd}
\usepackage{dcolumn}
\usepackage{graphicx}
\usepackage{subfigure}
\usepackage{float}
\usepackage[dvipsnames]{xcolor}
\usepackage[colorlinks,bookmarks=false,citecolor=blue,linkcolor=red,urlcolor=blue]{hyperref}
\begin{document}
\title{Self-bound  Doubly-Dipolar  Bose-Einstein condensates}

\author{Chinmayee Mishra}
      \affiliation{Indian Institute of Science Education and Research, Pune 411 008, India}   
      \author{Luis Santos}
      \affiliation{Institut f\"ur Theoretische Physik, Leibniz Universit\"at Hannover, Appelstrasse 2, DE-30167 Hannover, Germany}                    
\author{Rejish Nath}
       \affiliation{Indian Institute of Science Education and Research, Pune 411 008, India}
\date{\today}

\begin{abstract}
We analyze the physics of self-bound droplets in a doubly dipolar Bose-Einstein condensate~(DDBEC) composed by particles with both electric and magnetic dipole moments. 
Using the particularly relevant case of dysprosium, we show that the anisotropy of the doubly-dipolar interaction potential is highly versatile and non-trivial, depending critically on the relative orientation and strength between the two dipole moments.  This opens novel possibilities for exploring intriguing quantum many-body physics. Interestingly, by varying the angle between the two dipoles we find a dimensional crossover from quasi one-dimensional to quasi two-dimensional self-bound droplets. 
This opens a so far unique scenario in condensate physics, in which a dimensional crossover is solely driven by interactions in the absence of any confinement. 
\end{abstract}
\maketitle

Ultracold atoms are in the forefront for probing quantum many-body physics  \cite{blo08,pol11,gro17} due to the ability to manipulate them using external fields and engineering both short \cite{chi10} and long-range interatomic potentials \cite{sch12,tan18}.  Among them, dipolar gases \cite{bar08,bar12,lah09} gained a lot of attention since the first realization of the dipolar Bose-Einstein condensation (DBEC) of chromium atoms \cite{gri05, bea08}. The quest for systems with higher dipole moments paved the way for polar molecules \cite{gad16},  ultra cold Rydberg atoms \cite{saf10}, erbium \cite{aik12} and dysprosium (Dy) \cite{min11} BECs. For the first two cases, it is the electric dipole moment whereas in atomic condensates it is magnetic in nature. Due to the potential applications in quantum simulation \cite{mic06}, computing \cite{kar16}, tests of fundamental symmetries \cite{hud11} and tuning of collisions and chemical reactions \cite{abr07}, a growing interest  is focusing on
  particles possessing both electric and magnetic dipole moments \cite{lep18, pio10,tom14,ree17,rva17}. Although these studies have been mostly restricted to paramagnetic polar molecules \cite{pas13,bar14,khr14,gut18}, a recent proposal has discussed the feasibility of realizing an electric, in addition to the magnetic, dipole moment in Dy atoms \cite{lep18}. The latter is particularly important step due to the experimental feasibilities in atom-based setups, and opens intriguing questions about the effect of {\em doubly} dipole-dipole interactions (DDDIs) in quantum gases. 

Due to its anisotropic and long-range nature,  dipole-dipole interaction (DDI) results in novel phenomena in DBECs \cite{bar08,bar12,lah09}, including anisotropic superfluidity \cite{tic11,bis12,wen18}, roton-like excitations \cite{cho18, pet18}, quantum droplets \cite{Kad16, fer16,cho16,sch16} and dipolar supersolid states \cite{lta18, bot19, cho19,nat19,guo19,tan19}. The last two phenomena are rooted in the idea of quantum stabilization, the stabilization of a collapsing BEC due to the effective repulsion introduced by quantum fluctuations \cite{pet15}. Such a phenomenon is also responsible for the recently observed droplets in binary condensates \cite{cab18, sem18}. 

In this letter, we discuss the novel physics of condensates of particles with both magnetic and electric dipole moments. Focusing on the particularly relevant case of 
Dy atoms, we show that the combined DDDI exhibits a non-trivial anisotropy, which may be readily controlled by means of the relative orientation and strength of both dipole moments.
This nontrivial anisotropy results in an intriguing physics for doubly-dipolar quantum droplets. Whereas single-dipolar droplets are always elongated along the direction of the dipole moment, 
we show that a doubly-dipolar condensate allows for a dimensional crossover from a quasi one-dimensional (Q1D)  to a quasi two-dimensional (Q2D) regime as a function of the angle between the 
external electric and magnetic fields. This opens a so far unique scenario in condensate physics, in which a dimensional crossover is solely driven by interactions and completely unsupported by any confining potentials. 

\begin{figure}
\centering
\includegraphics[width= 1.\columnwidth]{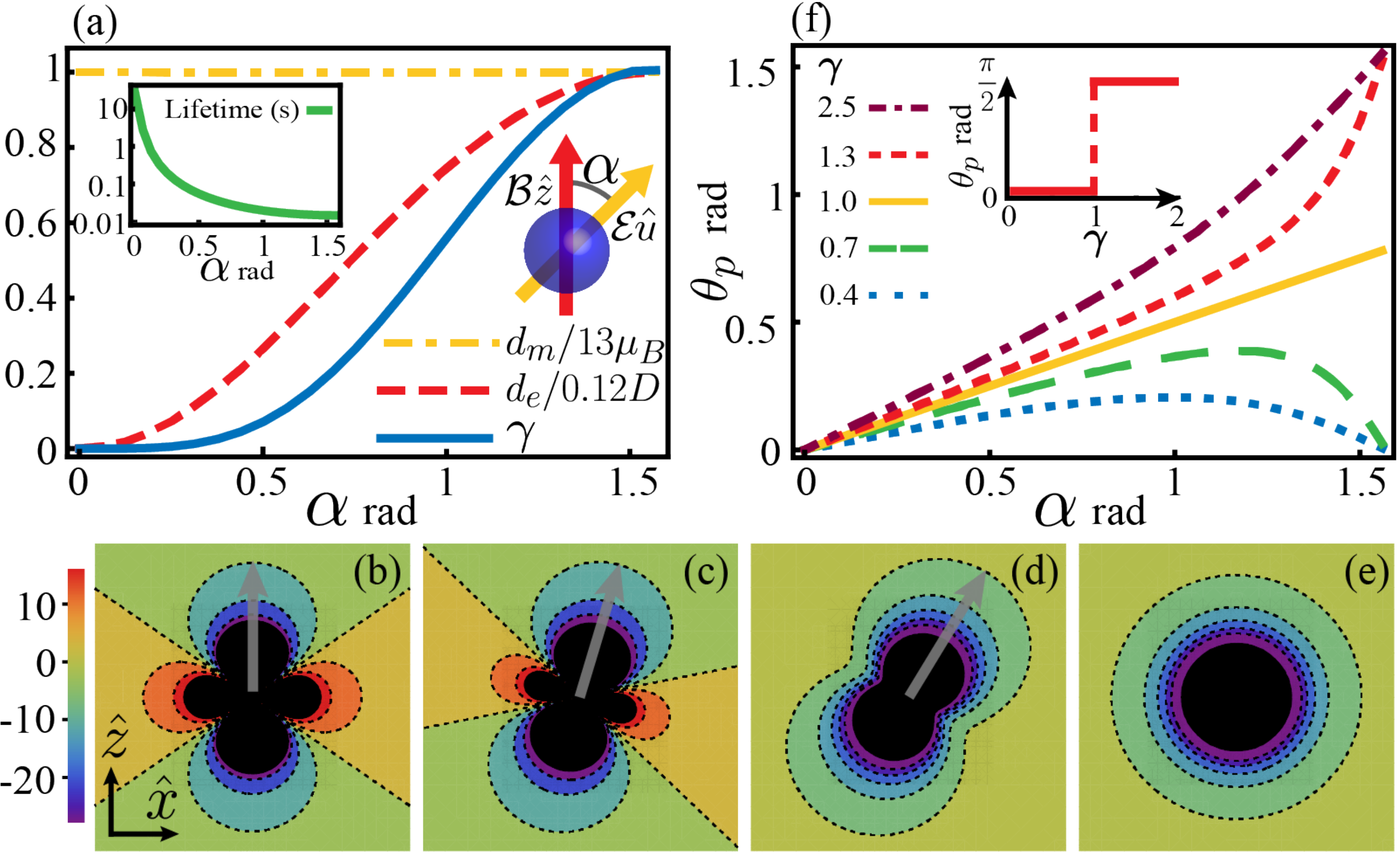}
\caption{\small{(color online). (a) The dipole moments  $d_m$ and $d_e$ of a Dy atom in the stretched state $|S\rangle$ as a function of $\alpha$  for $\mathcal{B}=100$ G and $\mathcal{E}=2.68$ kV/cm. With these field strengths the maximum values of $d_m$ and $d_e$ attained are 13$\mu_B$ and 0.12 Debye, respectively. The solid line shows $\gamma=g_e/g_m$. The inset shows the lifetime of the state $|S\rangle$. The 2D potential, $V_d^{y=0}({\bm r})$ (in arbitrary units), taking $d_m$ and $d_e$ from (a) for (b) $\alpha=0$, (c) $\alpha=1$ rad, $\alpha=1.3$ rad and $\alpha=\pi/2$ rad. For $\alpha>\alpha_a=1.239$ rad, $V_d^{y=0}({\bm r})$ is purely attractive, see (d) and (e). The thick arrow shows the effective polarization direction. (f) Polarization angle $\theta_p$ vs $\alpha$ for different $\gamma$; the inset shows $\theta_p$ as a function of $\gamma$ for $\alpha=\pi/2$.}}
\label{fig:1} 
\end{figure}

{\em Setup}.--- We consider a gas of $N$ bosons of mass $M$ with both electric and magnetic dipole moment, which are polarized, respectively,  by an external electric and magnetic field, 
which form an angle $\alpha$ between them. At very low temperatures, the system is described in mean field, by a nonlocal Gross-Pitaevskii equation (NLGPE): $i\hbar\dot{\psi}(\bm r,t)=\mathcal H\psi(\bm r,t)$, where 
\begin{eqnarray}
\mathcal H=\dfrac{-\hbar^2\nabla^2}{2M}+\int d^3r'\psi(\bm r', t)V(\bm r-\bm r')\psi(\bm r', t)
\label{nlgpe}
\end{eqnarray}
where $V(\bm r)=g\delta(\bm r)+V_d(\bm r)$ is the interaction potential. The parameter $g=4\pi\hbar^2 a_s N/M$ determines the contact interaction strength, with $a_s$ being the $s$-wave scattering length. The total dipole-dipole potential is, 
\begin{equation}
V_d(\bm r)=\frac{g_m}{r^3}\left[(1-3\cos^2\theta_m)+\gamma(1-3\cos^2\theta_e)\right]
\label{dddi}
\end{equation}
 where $\theta_m$ ($\theta_e$) is the angle formed by the magnetic (electric) dipole vector with $\bm{r}$,  $g_m=N\mu_0d_m^2/4\pi$ ($g_e=Nd_e^2/4\pi\epsilon_0$) is the magnetic (electric) DDI strength with $\mu_0$ ($\epsilon_0$) being the vacuum permeability (permittivity) and $d_m$ ($d_e$) is the magnetic (electric) dipole-moment. 
 
{\em Doubly-dipolar dysprosium Atoms}.---  As recently discussed in Ref.~\cite{lep18}, an electric dipole-moment can be induced in a Dy atom, in addition to the permanent magnetic moment. 
We consider a pair of quasi-degenerate states: $|a\rangle$ (odd parity) and $|b\rangle$ (even parity), with total angular momenta $\{J_a=10, J_b=9\}$, and energies $\{E_a=17513.33 \  \rm{cm}^{-1}, E_b=17514.50 \ \rm{cm}^{-1}\}$ \cite{lep18}.  A uniform magnetic field $\bm{B}=\mathcal B \hat{\bm{z}}$ along the $z$ axis which sets the quantization axis, splits the degeneracy of the states $|a\rangle$ and $|b\rangle$. A uniform electric field, $\bm{E}=\mathcal E \hat{\bm{u}}$ mixes the Zeeman sublevels $\{|M_a=-J_a\rangle, ... , |+J_a\rangle, |M_b=-J_b\rangle, ... , |+J_b\rangle\}$ of $|a\rangle$ and $|b\rangle$, thereby inducing an electric dipole moment along $\hat{\bm{u}}$ forming an angle $\alpha$ with the $z$-axis [see inset of Fig. \ref{fig:1}]. The electric field strength is such that the lowest eigenstate of the atom is $|S\rangle= c_0|M_a=-10\rangle+\sum_i' c_i|i\rangle$ with $\sum_i'|c_i|^2/|c_0|^2\ll 1$, where the sum $\sum_i'$ is over all the magnetic sublevels except $M_a=-
 10$. Within the electric-dipole approximation, the line widths of the states are $\Gamma_a\approx  0$ (metastable) and $\Gamma_b=2.98 \times 10^4$ s$^{-1}$ for the states $|a\rangle$ and $|b\rangle$ respectively. Thus, the lifetime of the stretched state $|S\rangle$ is $(n_b\Gamma_b)^{-1}$, determined by the total population ($n_b$) in $\{|M_b\rangle\}$ sublevels. An atom in $|S\rangle$ possesses a permanent magnetic dipole moment of $ {\bm d}_m=d_m \hat{{\bm z}}$ ($d_m\approx 13\mu_B$), and a tunable, induced electric dipole moment, ${\bm d}_e=d_e\hat{\bm{u}}$. This is illustrated in Fig. \ref{fig:1}(a), where we depict the dipole moments for realistic field strengths $\mathcal{B}=100$ G and $\mathcal{E}=2.68$ kV/cm \cite{mai15, nik10} as a function of $\alpha$ as well as the lifetime of the corresponding $|S\rangle$ \footnote{See supplemental material for the calculation details of ${\bm d}_{m,e}$, the derivation of $\theta_p$, the $N$-dependent axial and planar widths of the Q1D cigar and Q2D pancake droplets and the estimation of real and imaginary part of $\Delta\mu$.}. A BEC of 
 Dy atoms in state $|S\rangle$ is hence described by Eq.~\eqref{nlgpe}. Note that, the relative DDI strength, $\gamma=g_e/g_m$, depends explicitly on $\alpha$ [see Fig. \ref{fig:1}(a)] and the electric field strength $\mathcal E$. When $\alpha=0$ ($\gamma=0$) we have the case of a usual DBEC.
 
{\em Doubly dipole potential}.--- Considering the two dipoles oriented on the $xz$ plane, $V_d(\bm r)$ is always repulsive along the $y$-axis, whereas it exhibits a non-trivial anisotropy on the $xz$-plane. Figures \ref{fig:1}(b)-(e) show the 2D $xz$ potential [$V_d^{y=0}({\bm r})$] as a function of $\alpha$ for the Dy dipole-moments shown in Fig. \ref{fig:1}(a). It becomes purely attractive [Figs. \ref{fig:1}(d) and (e)] when $\alpha>\alpha_a=\frac{1}{2}\cos^{-1}\left[(-4\gamma^2+\gamma-4)/9\gamma\right]$,  but remains anisotropic except when $\alpha=\pi/2$ and $\gamma=1$. In the latter case we have $V_d^{y=0}({\bm r})=-1/r^3$, isotropic in the $xz$ plane. Nevertheless, we identify an effective polarization axis, depicted by arrows in Fig. \ref{fig:1}(b)-(d), along the direction in which the interaction between the atoms is maximally attractive. The polarization axis is determined by the angle \cite{Note1},
\begin{equation}
\theta_p(\alpha,\gamma)=\cos^{-1}\left[\frac{1}{\sqrt{2}}\sqrt{1+\frac{1+\gamma\cos2\alpha}{\sqrt{1+\gamma^2+2\gamma\cos2\alpha}}}\right],
\label{tp1}
\end{equation}
 with the 
 $z$-axis. The polarization angle $\theta_p$ exhibits as a function of $\alpha$ a monotonous~(non-monotonous) behavior for $\gamma\geq 1$~($\gamma<1$). A linear relation, $\theta_p=\alpha/2$ holds for $\gamma=1$ [see Fig. \ref{fig:1}(f)].   When $\alpha=\pi/2$ and $\gamma=1$, $\theta_p$ is not uniquely defined due to the radial symmetry of the $xz$-interactions. As a consequence, $\theta_p$ changes abruptly from zero to $\pi/2$ across $\gamma=1$ for $\alpha=\pi/2$ [inset of Fig. \ref{fig:1}(f)]. The angle $\theta_p$ plays a key role in the physics of quantum droplets discussed below.

{\em Homogeneous DDBEC}.--- The dispersion law of elementary excitations of a uniform DDBEC with density $n_0$ is 
\begin{eqnarray}
\varepsilon_{\bm k} =\sqrt{\dfrac{\hbar^2 k^2}{2M}\left(\dfrac{\hbar^2 k^2}{2M}+2g_m n_0\left[\beta+\mathcal F(\theta_k,\phi_k,\alpha)\right]\right)}
\label{disp3d}
\end{eqnarray}
where $\beta=g/g_m$ and
\begin{eqnarray}
\mathcal F(\theta_k,\phi_k,\alpha)&=&\frac{4\pi\gamma}{3}\left[3\left(\cos\alpha\cos\theta_k+\sin\alpha\sin\theta_k\cos\phi_k\right)^2-1\right] \nonumber \\
&&+\frac{4\pi}{3}(3\cos^2\theta_k-1).
 \label{fd}
\end{eqnarray}
with $\theta_k$ and $\phi_k$ the angular coordinates in momentum space. 
The long-wavelength excitations (phonons) are: $\varepsilon_{\bm k\to 0}=c(\theta_k,\phi_k)\hbar k$ with the angle-dependent sound velocity
\begin{equation}
c(\theta_k,\phi_k)=\left[g_m n_0\left(\beta+\mathcal F(\theta_k,\phi_k,\alpha)\right)/M\right]^{1/2}.
\end{equation} 
The stiffer phonons propagate along the effective polarization axis whereas the softer ones are perpendicular to it. The phonons along the $y$-axis are always softer which sets the stability criteria for the condensate $c^2(\pi/2,\pi/2)=\frac{g_mn_0}{M}\left[\beta-\frac{4\pi}{3}(1+\gamma)\right]>0$. Thus, a homogeneous DDBEC becomes unstable against local collapses if $\beta<\frac{4\pi}{3}(1+\gamma)$. 


 {\em Self-bound droplet}.---Although the mean-field instability criterion has no explicit dependence on $\alpha$, the effect of quantum fluctuations depends critically on it. Using the dispersion~\eqref{disp3d}, we obtain the beyond mean field, Lee-Huang-Yang (LHY), correction to the chemical potential ($\Delta\mu$) \cite{sch06,wac16, bis16,wac16-2,sai16}:
 \begin{eqnarray}
\Delta\mu=\frac{1}{3\pi^3N}\left(\frac{M}{\hbar^2}\right)^{3/2}n_0^{3/2}g_m^{5/2}  \int d\Omega_k\left[\beta+\mathcal F(\theta_k,\phi_k,\alpha)\right]^{\frac{5}{2}}
\label{lhy}
\end{eqnarray}
 where $\int d\Omega_k=\int_0^{2\pi}d\phi_k\int_0^\pi d\theta_k\sin\theta_k$. The correction, $\Delta\mu$ becomes complex when $\beta<\frac{4\pi}{3}(1+\gamma)$. The real part of $\Delta\mu$ is dominated by hard modes, whereas the unstable low-momentum excitations determine the imaginary part. Not very deep in the instability regime, ${\rm Im}[\Delta\mu]/{\rm Re}[\Delta\mu] \ll 1$ and ${\rm Im}[\Delta\mu]$ can be disregarded \cite{Note1}. Note also that for a finite size condensate or a droplet, ${\rm Im}[\Delta\mu]$ is further suppressed by a low-momentum cut-off \cite{Note1,bis16, sai16, wac16}.

 Due to the $n_0^{3/2}$ dependence, the repulsive LHY correction becomes significant at high densities, 
 and halts the condensate collapse, stabilizing it into a quantum droplet. The LHY correction can be incorporated to the 
 NLGPE using local density approximation [$n_0\to n(\bm r, t)$] \cite{lim11, wac16, bis16,wac16-2,sai16, bai16, bai17}], leading to the generalized equation:
 \begin{equation}
 i\hbar\dot{\psi}(\bm r,t)=\left(\mathcal H+\Delta\mu\left[n(\bm r, t)\right]\right)\psi(\bm r,t).
 \label{ggpe}
 \end{equation}
 Similar generalized GPE equations have been employed in singly-dipolar BECs  and binary mixtures \cite{cho16, sch16, bot19, cho19, nat19, pet15, cab18, sem18}, providing a good qualitative and to a large extent quantitative agreement with experiments. A better quantitative picture  may however require going beyond the local-density approximation and/or the use of involved quantum Monte Carlo calculations \cite{mac16,cin17,cint17, sta18, cik19,bott19}. 
 
\begin{figure}
\centering
\includegraphics[width= 1.\columnwidth]{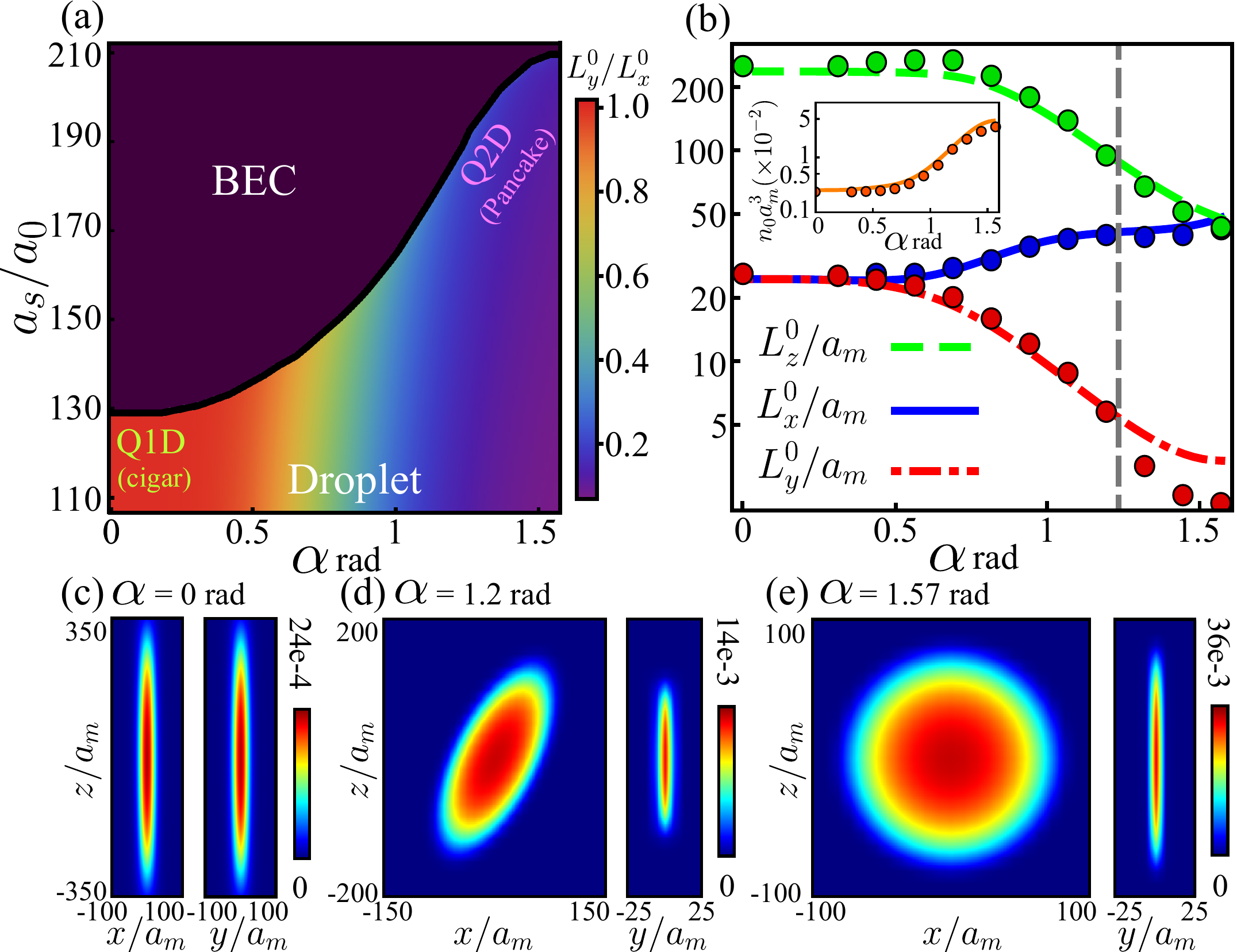}
\caption{\small{(color online). (a) Stable/unstable region for a self-bound (Gaussian) droplet of Dy atoms in the state $|S\rangle$, for $\mathcal{B}=100$ G, $\mathcal{E}=2.68$ kV/cm and $N=2000$, as function of the scattering length $a$ and $\alpha$. The stable regime is partitioned using $L_y^0/L_x^0$. (b) Equilibrium widths from the Gaussian calculations (solid lines), and the 3D numerical simulations of Eq. (\ref{ggpe}) (filled circles) as a function of $\alpha$ for $a=110 a_0$ where $a_0$ is the Bohr radius and other parameters as in (a). The inset shows the peak density of the droplet. The angle $\alpha_a$ is shown by a dashed vertical line. The widths are scaled by the length $a_m=\mu_0 m d_m^2/12\pi\hbar^2\simeq 222a_0$. (c)-(e) show the density plots for the droplet ground state from 3D numerics for different $\alpha$, indicating a structural crossover from cigar to pancake-shaped droplet.}}
\label{fig:2} 
\end{figure}

 We introduce at this point a  time-dependent variational Gaussian ansatz that is particularly helpful to understand the key properties of doubly-dipolar droplets:
\begin{eqnarray}
\psi({\bm r}, t)&=&\dfrac{1}{\pi^{3/4}\sqrt{L_xL_yL_z}}\exp\left[-\dfrac{x'^2}{2L_x^2}-\dfrac{y^2}{2L_y^2}-\dfrac{z'^2}{2L_z^2}+ \right.\nonumber \\
&&\left. ix'^2\beta_{x}+iy^2\beta_y+iz'^2\beta_{z}+ix'z'\beta_{xz}\right],
\label{vg0}
\end{eqnarray}
where $x'=x\cos\theta-z\sin\theta$, $z'=x\sin\theta+z\cos\theta$, and $\{L_{x, y, z}, \beta_{x, y, z}, \theta\}$ are variational parameters. In particular, $L_{x, y, z}$ is the droplet width 
along the $\{x', y, z'\}$ direction and $\theta$ determines its orientation in the $xz$-plane. Employing Eq. (\ref{vg0}) in the Lagrangian density describing a DDBEC, 
\begin{widetext}
\begin{eqnarray}
\mathcal{L}= \dfrac{i\hbar}{2}\left(\psi\dot\psi^*-\dot\psi\psi^*\right)+\dfrac{\hbar^2}{2m}|\nabla\psi|^2+\dfrac{g}{2}|\psi|^4+\dfrac{1}{2}|\psi|^2\!\! \int \!d^3r'V_d(r-r')|\psi(r')|^2+
\dfrac{2M^{3/2}}{15\pi^3\hbar^3N}
g_m^{5/2}|\psi|^5\!\! \int \!\! d\Omega_k[\beta+\mathcal{F}(\theta_k,\phi_k,\alpha)]^{5/2}
\end{eqnarray}
\end{widetext}
we obtain the Lagrangian $L=\int d^3r \mathcal{L}$ and the Euler-Lagrange equations of motion, $\dfrac{d}{dt}\left(\dfrac{\partial L}{\partial \dot{Q}}\right)-\dfrac{\partial L}{\partial Q}=0$, with 
$Q\in \{L_{x, y, z}, \beta_{x, y, z}, \theta\}$. The latter provides an effective potential describing the evolution of the dynamical variables, and by minimizing it we evaluate the equilibrium values for the widths ($L_i^0$) and the orientation ($\theta^0$)  \cite{Note1}. 
We obtain $\theta^0=\theta_p$, as given in Eq. (\ref{tp1}).  
A finite $L_i^0$ indicates a stable self-bound droplet. In  Fig. \ref{fig:2}(a) we show the stable/unstable region (evaluated using the Gaussian ansatz) for a self-bound droplet in a Dy DDBEC as a function of $a_s$ and $\alpha$, for $N=2000$ and the dipole moments given in Fig. \ref{fig:1}(a).  For a large-enough $a_s$, the repulsive short-range interactions overcome the attractive part of the DDDI, resulting in a mean-field stable BEC. If $a_s$ is below an $\alpha$-dependent critical value ($a_{cr}$), the attractive mean-field interaction is balanced by the repulsive 
LHY correction resulting in a self bound droplet. Note that $a_{cr}$ is larger for higher $\alpha$ since 
the attractive region in $V_d^{y=0}({\bm r})$ increases~[see Figs. \ref{fig:1}(b)-(e)]. 

The properties of the self-bound droplet are crucially determined by $\alpha$. In particular, when $\alpha=0$, the droplet is radially symmetric $\left(L_x^0=L_y^0\right)$, cigar-shaped with the long axis parallel to the $z$-axis [see Figs. \ref{fig:2}(b) and \ref{fig:2}(c)]. This is so far the only scenario reported in DBECs \cite{Kad16, fer16,cho16,sch16}. When $\alpha$ increases, the effective polarization axis changes and the droplet gets tilted in the $xz$ plane by an angle $\theta_p$ [see Fig. \ref{fig:2}(d)]. The tilting is accompanied by a structural deformation due to the change in the anisotropic properties of the DDDI with $\alpha$. $L_z^0$ and $L_y^0$ decrease whereas $L_x^0$ increases with $\alpha$ [see Fig. \ref{fig:2}(b)]. For 
$\alpha_a<\alpha<\pi/2$~[with $\alpha_a\simeq 1.2\mathrm{rad}$ in Fig. \ref{fig:2}(b)]  the droplet acquires an anisotropic~($L_x^0\neq L_z^0$) pancake shape on the $xz$-plane.  A pancake condensate with $L_x^0=L_z^0$ results for $\alpha=\pi/2$ ($\gamma=1$) [see Fig. \ref{fig:2}(e)]. To capture the crossover from a cigar- to a pancake-shape droplet as a function of $\alpha$, the droplet region in Fig. \ref{fig:2}(a) is partitioned using the ratio $L_y^0/L_x^0$. For the parameters considered here, the pancake droplet is denser than the cigar one [see the inset of Fig. \ref{fig:2}(b) for the peak density of the droplet]. 
Although the actual condensate profile, which we obtain by imaginary time evolution of Eq.~\eqref{ggpe} is not Gaussian, the evaluated widths of the DDBEC are in excellent 
quantitative agreement with the Gaussian results [see Fig. \ref{fig:2}(b)].

\begin{figure}
\centering
\includegraphics[width= 1.\columnwidth]{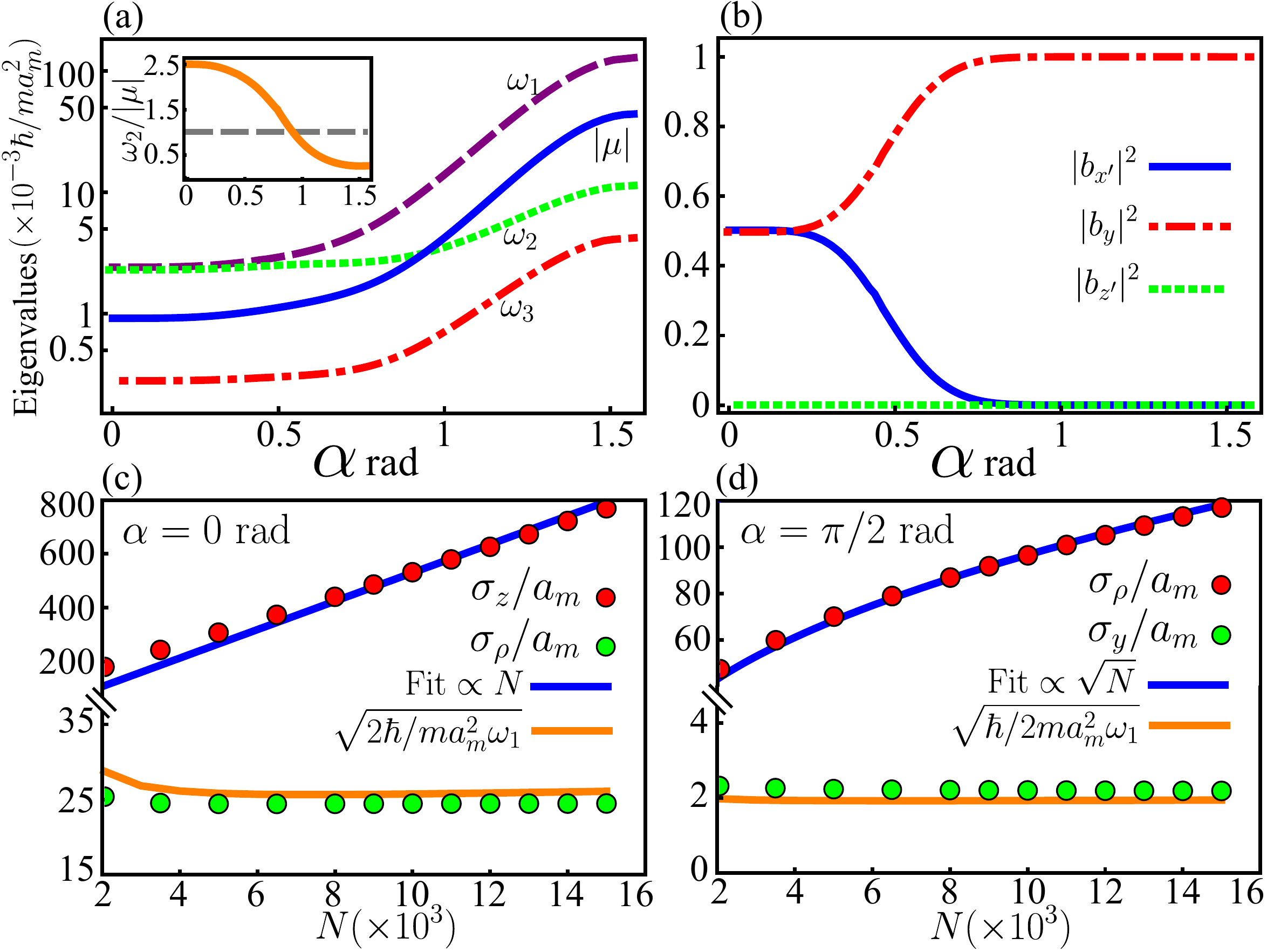}
\caption{\small{(color online). (a) Excitation frequencies of the Dy droplet obtained from the Gaussian variational calculations as a function of $\alpha$ for the same parameters as in Fig. \ref{fig:2}(b). The inset shows $\omega_2/|\mu|$ as a function of $\alpha$. (b) The eigenvector $\sum_{j=x',y,z'} b_j \vec e_j$ corresponding to 
$\omega_1$ as a function of $\alpha$. It changes from being a breathing mode on the $xy$-plane for $\alpha=0$ to a pure $y$-oscillation as $\alpha$ approaches $\pi/2$. The $N$-dependence of the droplet widths at $\alpha=0$ and $\alpha=\pi/2$ is shown in (c) and (d) respectively.}}
\label{fig:3} 
\end{figure}

{\em Collective Excitations}.--- The low-lying excitations of the self-bound droplet are obtained by linearizing the equations of motion for $Q$ around the potential minimum. 
The three modes associated with small oscillations of the droplet widths are shown in Fig. \ref{fig:3}(a), where they are compared with the chemical potential $\mu$ of the droplet. 
At $\alpha=0$ (cigar-shaped droplet) the two transverse modes correspond to the breathing ($\omega_1$) and the quadrupole motion ($\omega_2$) of the droplet in the $xy$ plane whereas $\omega_3$ represents the axial motion. Interestingly, we have $\omega_1>\omega_2$ with $\omega_2\simeq 2.5|\mu|$  [see inset of Fig. \ref{fig:3}(a)] which confirms that the droplet is Q1D in nature. At $\alpha=\pi/2$ (pancake droplet on the $xz$ plane) the transverse excitation ($\omega_1$) has a purely $y$-character [see Fig. \ref{fig:3}(b)] with $\omega_1\simeq 2.8|\mu|$, confirming  the Q2D character of the droplet. The modes having frequencies $\omega_2$ and $\omega_3$ become respectively breathing and quadrupole modes in the $xz$ plane with $\omega_2>\omega_3$. 
Excitations reveal that more than just a structural crossover, there exists an actual dimensional crossover from Q1D to Q2D regime as a function of $\alpha$. Alternatively, as evident from Fig. \ref{fig:1}(f), the dimensional crossover can be observed by varying the electric field strength (or equivalently $\gamma$) keeping $\alpha=\pi/2$. In this case, when $\gamma\ll 1$ we have a Q1D droplet along the $z$-axis, as $\gamma\to 1$ the droplet becomes a Q2D, and for $\gamma\gg 1$ again a Q1D droplet but along the $x$-axis.

{\em $N$-dependence}.--- Finally we analyze the dependence of the droplet widths with $N$, which provide a good witness of the dimensional crossover. 
Deep in the Q1D ($\alpha\sim 0$) and  Q2D ($\alpha \sim \pi/2$) regimes the transverse widths are almost independent of $N$ and they can be estimated from the transverse modes by 
analogy to low-dimensional trapped condensates. For the Q1D droplet we have $L_x=L_y=\sqrt{2\hbar/m\omega_1}$ which agrees excellently with the numerical results of $\sigma_{\rho}=\sqrt{\int d^3r(x^2+y^2)|\psi(r)|^2}$ [see Fig. \ref{fig:3}(c)]. For the Q2D case, we have $L_y=\sqrt{\hbar/m\omega_1}=\sqrt{2}\sigma_y$, where $\sigma_{y}=\sqrt{\int d^3ry^2|\psi(r)|^2}$ [see Fig. \ref{fig:3}(d)]. To estimate the $N$-dependence of axial ($\alpha=0$) and the in-plane ($\alpha=\pi/2$) widths we employ a hybrid Gaussian-TF ansatz in Eq. (\ref{ggpe}) \cite{Note1}. After taking the appropriate limits: $L_{x,y}/R_z\ll 1$ (Q1D) and $L_y\ll R_{\perp}$ where $R_{z,\perp}$ are the respective TF radii, we get 
\begin{eqnarray}
&&R_z\simeq\dfrac{81a_m^3 N}{25\pi^7L_x^2\left(4\pi/3-\beta\right)^2}\left[\int d\Omega_k\left[\beta+\mathcal{F}(\theta_k,\phi_k,0)\right]^{5/2}\right]^2, \\
&&R_\perp\simeq\dfrac{9a_m^{3/2}}{2\pi^{13/4}}\sqrt{\dfrac{N}{5L_y}}\dfrac{\int d\Omega_k[\beta+\mathcal{F}(\theta_k,\phi_k,\pi/2)]^{5/2}}{\left(8\pi/3-\beta\right)}.
\end{eqnarray}
respectively for Q1D and Q2D droplets. These $N$-dependences are in excellent agreement with the results of our three-dimensional simulations of the generalized NLGPE, 
as shown in Figs. \ref{fig:3}(c) and \ref{fig:3}(d).

{\em Conclusions}.--- The doubly-dipolar potential exhibits non-trivial, yet versatile anisotropic properties, which leads to new features in self-bound quantum droplets. In particular, 
as a function of the relative orientation of the electric and magnetic dipole moments, the droplet undergoes a structural crossover from a cigar to a pancake shape. A detailed analysis on the collective excitations and droplet widths reveal that, the structural crossover is associated with a dimensional crossover from Q1D to Q2D regime. This offers a unique scenario in BECs in which an interaction-induced dimensional crossover can be studied in the absence of any confining potentials. Our results open up exciting possibilities for novel quantum phenomena 
both in lattice models and in BECs. Since roton excitations in elongated DBECs occur for a momentum proportional to the inverse transverse confinement along the dipole 
direction, roton excitations in DDBECs are expected to present a peculiar nature, especially in elongated non-axisymmetric traps for $\alpha=\pi/2$, due to the competition between two 
different roton length scales. Moreover, Q2D droplets constitute an intriguing scenario for the study of vortex lattices, and hence of droplet superfluidity.

C. M. and R. N. acknowledge funding from the Indo-French Centre for the Promotion of Advanced Research, UKIERI-UGC Thematic Partnership No. IND/CONT/G/16-17/ 73UKIERI-UGC project, and Department of Science and Technology (DST), Government of India through the INSPIRE fellowship. L. S. thanks the support of the DFG (SFB 1227 DQ-mat and FOR2247).

\bibliography{libdbec.bib}

\newpage
\onecolumngrid
\newpage
{
	\center \bf \large 
	Supplemental Material for: \\
	Self-bound  Doubly-Dipolar  Bose-Einstein condensates
	\vspace*{0.1cm}\\ 
	\vspace*{0.0cm}
}

\begin{center}
	Chinmayee Mishra$^1$, Luis Santos$^2$ and Rejish Nath$^1$\\
	\vspace*{0.15cm}
	\small{\textit{$^1$Indian Institute of Science Education and Research, Pune 411 008, India}}\\
	\small{\textit{$^2$Institut f\"ur Theoretische Physik, Leibniz Universit\"at Hannover, Appelstrasse 2, DE-30167 Hannover, Germany}}\\
	\vspace*{0.25cm}
\end{center}

\section{Calculation of Dy dipole moments}
\begin{figure}[h]
	\begin{center}
		\includegraphics[width=0.5\columnwidth]{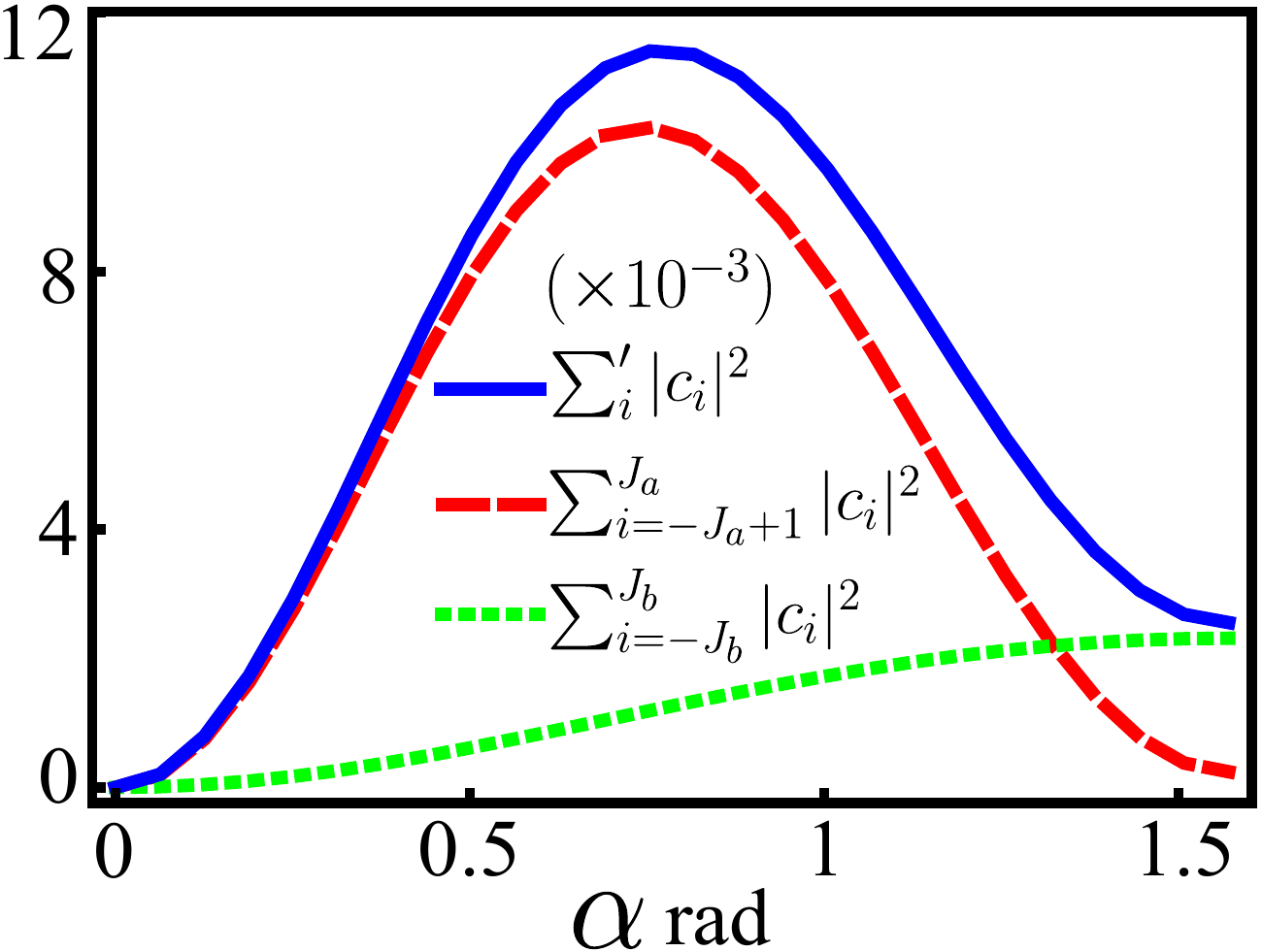}
		\caption{(Color online) For a fixed $\mathcal{B}=100$ G and $\mathcal{E}=2.68$ kV/cm, the probability of finding the stretched state $|S\rangle$ in all sub-levels other than $|M_a=-10\rangle$ is found to be extremely small for any $\alpha$ such that the maximum value of $\sum_i'|c_i|^2/|c_0|^2\approx0.0116$. The dashed lines show the contributions from the sublevels $\{|M_a\rangle\}$ and $\{|M_b\rangle\}$.}
		\label{fig:s1}
	\end{center}
\end{figure}

As mentioned in the main text, we consider a pair of quasi-degenerate energy levels with opposite parity and total angular momenta $J_a=10$ and $J_b=9$, in a Dysprosium atom \cite{lep18}. The Hamiltonian describing the atom restricted to the subspace of the two energy levels is $\hat{H}=E_a\sum_{M_a=-J_a}^{J_a}|M_a\rangle\langle M_a|+E_b\sum_{M_b=-J_b}^{J_b}|M_b\rangle\langle M_b|+\mu_B B (g_a M_a+g_b M_b)+\hat{H}_{stark}$, where $\{M_{a,b}\}$ are the magnetic sublevels, $g_a=1.3$ and $g_b=1.32$ are the Land\'e $g$ factors. $\hat{H}_{stark}$ accounts the electric field-atom interaction. The magnetic field splits the level into its Zeeman sub levels whereas the electric field admixes those having opposite parity. Finally, we are interested in the stretched state  $|S\rangle=c_0|M_a=-10\rangle+\sum_i'c_i|i\rangle$, where $\sum_i'$ is the sum over all the sublevels except $M_a=-10$ and the condition $\sum_i'|c_i|^2/|c_0|^2<<1$ is satisfied, see Fig. \ref{fig:s1}. The latter has two cont
ributions, one from the sublevels of the state $|a\rangle$ and second from that of $|b\rangle$. The contributions from the sublevels $\{|M_b\rangle\}$ determine the life time of the state $|S\rangle$. The field magnitudes are chosen such that irrespective of $\alpha$ the state $|S\rangle$ remains as the lowest eigenstate. 

The magnetic and electric dipole moments of Dy atom in the stretched state $|S\rangle$ is calculated as,
\begin{eqnarray}
	d_m&=&-\mu_B\left(g_a\sum_{M_a=-J_a}^{J_a}|c_{M_a}|^2M_a+g_b\sum_{M_b=-J_b}^{J_b}|c_{M_b}|^2M_b\right)\\
	d_e&=&-\dfrac{1}{\mathcal{E}}\sum_{M_a,M_b}c^*_{M_a}c_{M_b}\langle M_a|\hat{H}_{stark}|M_b\rangle+c.c.
\end{eqnarray}
where $c_i$ is the probability amplitude for finding the atom in state $|i\rangle$,
\begin{eqnarray}
	\langle M_a|\hat{H}_{stark}|M_b\rangle&=&-\sqrt{\dfrac{4\pi}{3(2J_a+1)}}\langle a||\hat{d}||b\rangle \mathcal{E}Y^*_{1,M_a-M_b}(\alpha,0)C^{J_aM_a}_{J_bM_b,1,M_a-M_b} \nonumber
\end{eqnarray}
where, $\langle a||\hat{d}||b\rangle=8.16 D$ is the reduced transition dipole moment, $Y_{l,m}(\theta,\phi)$ is the spherical harmonics and $C$s are the Clebsch-Gordon coefficients.

\section{Derivation of angles: $\theta_p(\alpha,\gamma)$ and $\alpha_a$}
\begin{figure}
	\centering
	\includegraphics[width= .7\columnwidth]{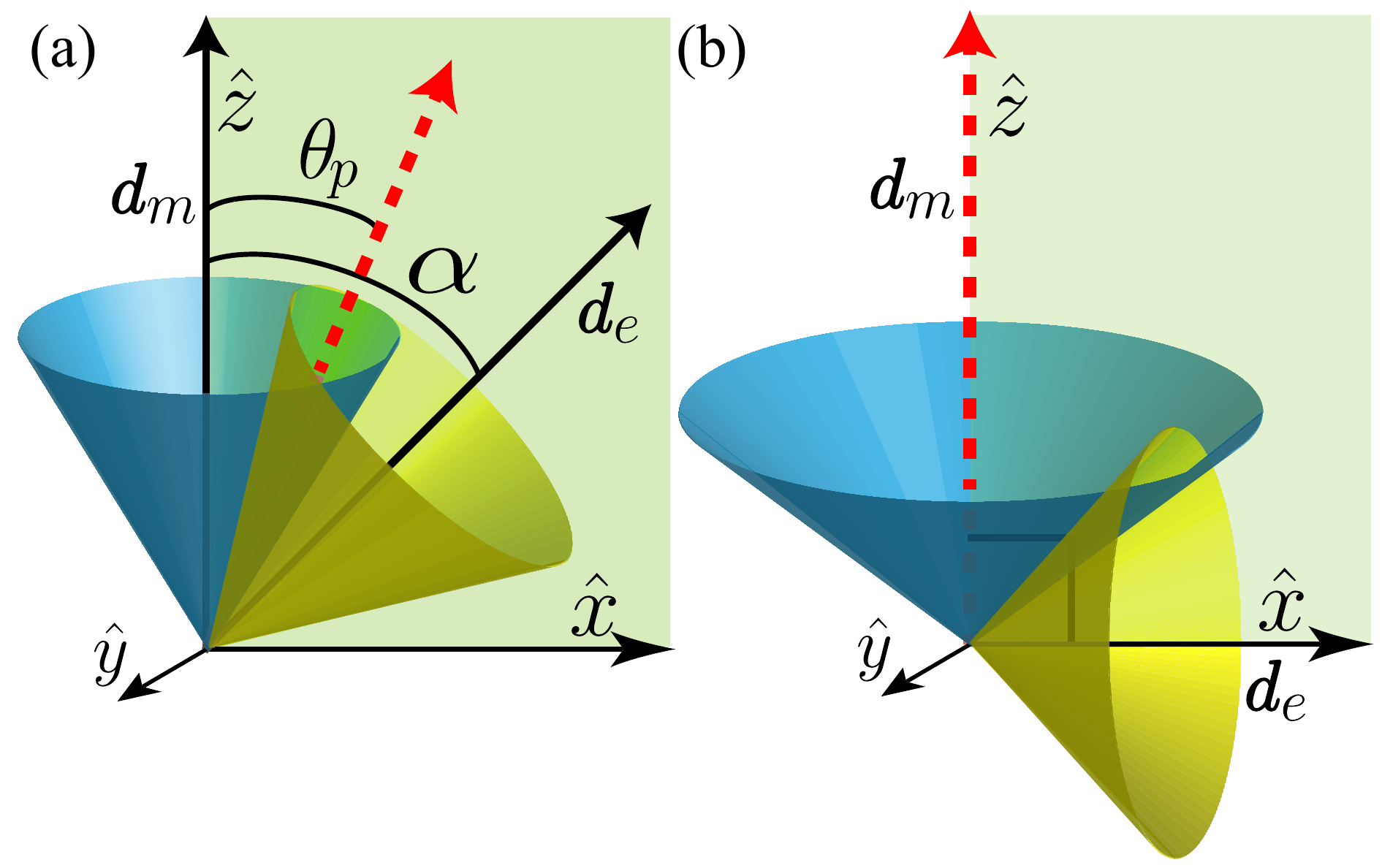}
	\caption{\small{(color online). Orientation of the electric and magnetic dipoles and the cones showing the volume where the dipole moments of the same kind experience attractive DDI. (a) The case in which both type of DDI have the same strength and the dipoles form an angle $\alpha$ between them. (b) The dipoles are mutually perpendicular to each other,  the effective polarization axis along the axis ($z$-axis) of the strongest dipole strength. If they are of same strength the polarization axis is undefined.}}
	\label{fig:s2} 
\end{figure}

The orientation of the electric and magnetic dipoles and the cones representing the regions in which the respective dipole moments in the second atom experiences attracting interactions are shown in Fig. \ref{fig:s2}(a)-(b). For Fig. \ref{fig:s2}(a), both the electric ($g_e$) and magnetic ($g_m$) DDI strengths are assumed to be of the same strength, and $\alpha$ is such that there is a significant overlap in the attractive cones. The angle $\theta_p$ representing the effective polarization axis lies at the centre of the overlapping region in the $xz$-plane, whereas in Fig. \ref{fig:s2}(b), in a different  scenario, both the dipoles are mutually perpendicular, the effective polarization axis is along the strongest dipole direction, i.e., along ${\bm d}_m$. In any case, the effective polarization axis lies always on the $xz$-plane. Below, we obtain a relation for $\theta_p$.

Using the spherical coordinates, the total dipole-dipole potential for two particles on the $xz$ plane ($y=0$) is,
\begin{eqnarray}
	V_d^{y=0}(r,\theta)&=&\dfrac{g_m(1+\gamma)}{r^3}\left[1-3\dfrac{\cos^2\theta+\gamma(\cos\theta\cos\alpha+\sin\alpha\sin\theta)^2}{1+\gamma}\right],
\end{eqnarray}
which is independent of $\phi$. The polarization angle, $\theta_p$ determines the direction along which the dipoles are maximally attractive, and is depicted in Figs. \ref{fig:s2}(a) and \ref{fig:s2}(b). It is obtained by minimizing $V_d^{y=0}(r,\theta)$ with respect to $\theta$ for a constant  $r$, and it gives us,
\begin{eqnarray}
	\theta_{p}=\cos^{-1}\left[\dfrac{1}{\sqrt{2}}\sqrt{1+\dfrac{1+\gamma\cos2\alpha}{\sqrt{1+\gamma^2+2\gamma\cos2\alpha}}}\right].
\end{eqnarray}

\begin{figure}
	\centering
	\includegraphics[width= 1.\columnwidth]{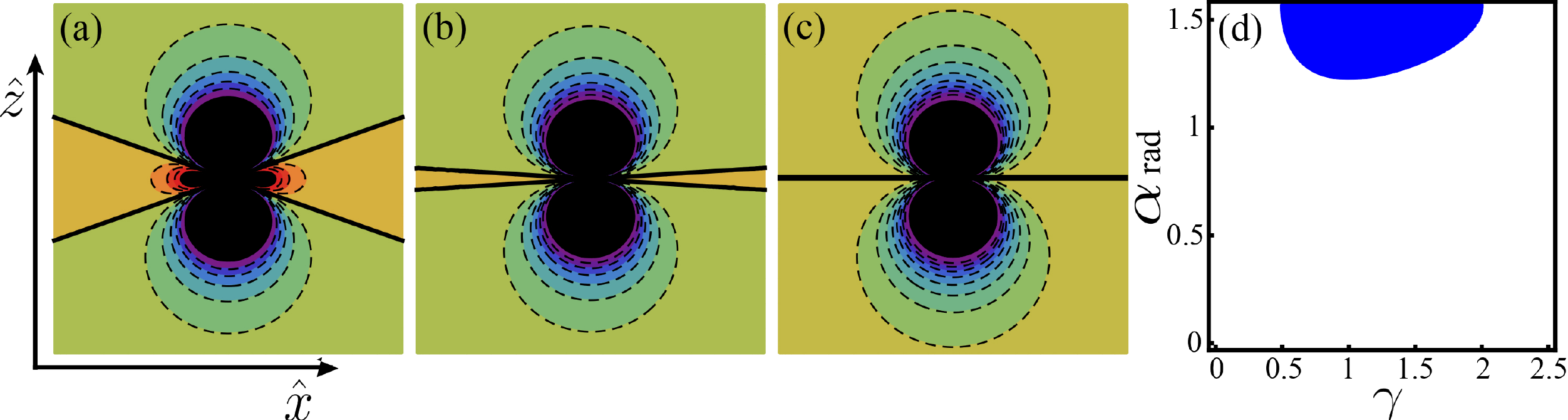}
	\caption{\small{(color online). (a)-(c) show the anisotropic interactions, $V_d^{y=0}({\bm r})$ for two dipoles in the $xz$ plane for $\alpha=\pi/2$ with (a) $\gamma=0.45$  (b) $\gamma=0.59$ and (c) $\gamma=0.5$. The repulsive region disappeared in (c). The shaded region in the $\alpha$-$\gamma$ plane shown in (d) is where $V_d^{y=0}({\bm r})$ is purely attractive.}}
	\label{fig:s3} 
\end{figure}

On the other hand, the angle $\alpha_a$ separates the regime between the purely attractive $xz$ potential ($\alpha>\alpha_a$) with an anisotropic $xz$ potential exhibiting both repulsive and attractive lobes.  The attractive and repulsive lobes in the potential are separated by two intersecting straight lines satisfying $V_d^{y=0}(r,\theta)=0$, see Fig. \ref{fig:s3}(a)-(c). Writing the equation for these straight lines as $z=m_{\pm}x$, where the slopes $m_\pm$ can be found from the condition $V_d^{y=0}(r,\theta)=0$ as,
\begin{eqnarray}
	m_{\pm}&=&-\dfrac{3\gamma\sin2\alpha\pm\sqrt{2(4-\gamma+4\gamma^2+9\gamma\cos2\alpha)}}{4+\gamma+3\gamma\cos2\alpha}.
\end{eqnarray}
When the $xz$ potential  becomes purely attractive the two lines merges into one, i.e., $m_+=m_-$, and we obtain
\begin{eqnarray}
	\alpha_a=\dfrac{1}{2}\cos^{-1}\left[\dfrac{\gamma-4(1+\gamma^2)}{9\gamma}\right]. 
\end{eqnarray}
The angle $\alpha_a$ is determined by the boundary of the shaded region in the $\alpha$ - $\gamma$ plane shown in Fig. \ref{fig:s3}(d).

\section{LHY correction $\Delta\mu$}
\begin{figure}[b]
	\centering
	\includegraphics[width= 0.8\columnwidth]{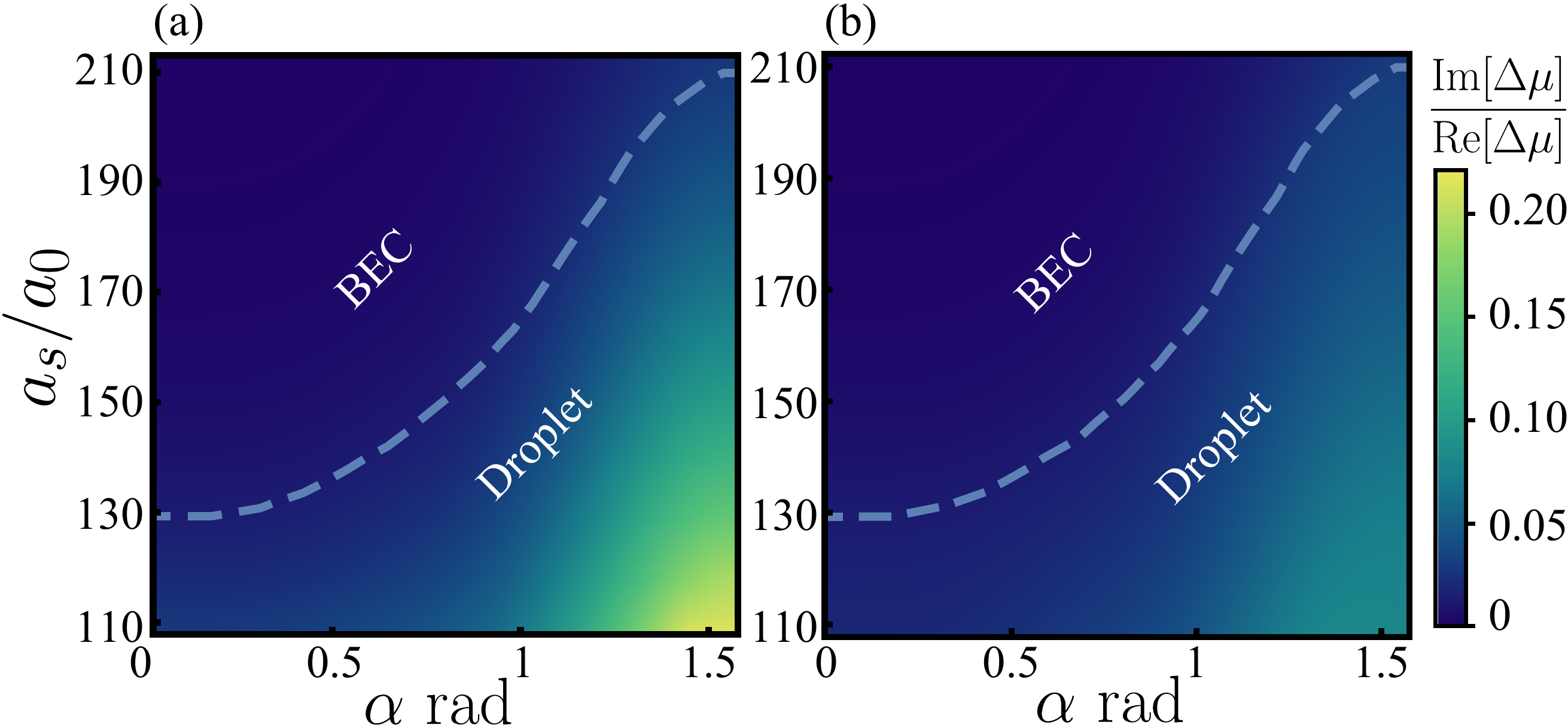}
	\caption{\small{(color online). (a) The ratio ${\rm Im}[\Delta\mu]/{\rm Re}[\Delta\mu]$ for a homogeneous condensate. }}
	\label{fig:s4} 
\end{figure}

The beyond mean field, Lee-Huang-Yang (LHY) correction to the chemical potential ($\Delta\mu$) for a homogeneous doubly-dipolar condensate is
\begin{eqnarray}
	\Delta \mu&=&\dfrac{5}{8\sqrt{2}\pi^3 N}\left(\dfrac{Mn_0}{\hbar^2}\right)^{3/2}g_m^{5/2}\int d\Omega_k\int_0^\infty k^2dk \left(k\sqrt{k^2+2[\beta+\mathcal{F}(\theta_k,\phi_k,\alpha)]}-k^2-\beta-\mathcal{F}(\theta_k,\phi_k,\alpha)+\dfrac{[\beta+\mathcal{F}(\theta_k,\phi_k,\alpha)]^2}{2k^2}\right)\label{def}
\end{eqnarray}
where $\int d\Omega_k=\int_0^{2\pi}d\phi_k\int_0^\pi d\theta_k\sin\theta_k$, $\mathcal{F}(\theta_k,\phi_k,\alpha)=\dfrac{4\pi}{3}(3\cos^2\theta_k-1)+\dfrac{4\pi\gamma}{3}[3(\cos\theta_k\cos\alpha+\sin\theta_k\cos\phi_k\sin\alpha)^2-1]$ and $\beta=g/g_m$. After doing the integration over $k$, we get the Eq. (7) in the main text. The LHY correction, $\Delta\mu$ becomes complex when $\beta<\frac{4\pi}{3}(1+\gamma)$. In order for the generalized NLGPE [Eq. (8) in the main text] to be valid, we require ${\rm Im}[\Delta\mu]/{\rm Re}[\Delta\mu] \ll 1$. Note that, the real part of $\Delta\mu$ is dominated by the hard modes in the Bogoliubov spectrum, whereas the imaginary part is determined by the unstable low momentum excitations.
The ration ${\rm Im}[\Delta\mu]/{\rm Re}[\Delta\mu]$ for a homogeneous BEC is shown in Fig. \ref{fig:s4}(a). As one can see that, in the vicinity of the boundary between a stable repulsive BEC and a droplet, ${\rm Im}[\Delta\mu]/{\rm Re}[\Delta\mu]$ is vanishingly small, and the ${\rm Im}[\Delta\mu]$ can be neglected. But, deep in the droplet regime, the imaginary part becomes significant, and the generalized NLGPE [Eq. (8) becomes invalid.

In a finite condensate or a droplet, the low-momentum excitations are absent, leading to a lower cutoff  ($k_c$) for the momentum integral in Eq. (\ref{def}). Let $C_x, C_y, C_z$ be the cut-offs along the generalised ellipsoid axes of our droplet. Then, the final integration is done over a region defined by \cite{},
\begin{eqnarray}
	\dfrac{(k_x\cos\theta_p-k_z\sin\theta_p)^2}{C_x^2}+\dfrac{k_y^2}{C_y^2}+\dfrac{(k_z\cos\theta_p+k_x\sin\theta_p)^2}{C_z^2}\geq 1. \nonumber
\end{eqnarray}
Thus, we have an angle dependent cut-off limit as,
\begin{eqnarray}
	k_c(\theta_k,\phi_k)=\dfrac{1}{\sqrt{\dfrac{(\sin^2\theta_k\cos\phi_k\cos\theta_p-\cos\theta_k\sin\theta_p)^2}{C_x^2}+\dfrac{\sin^2\theta_k\sin^2\phi_k}{C_y^2}+\dfrac{(\cos\theta_k\cos\theta_p+\sin\theta_k\cos\phi_k\sin\theta_p)^2}{C_z^2}}}
\end{eqnarray}
Implementing this cutoff in the $k$ integral of Eq. \eqref{def}, the (modified) LHY correction is obtained as,
\begin{eqnarray}
	\Delta\mu^c&=&\dfrac{1}{3\pi^3 N}\left(\dfrac{M}{\hbar^2}\right)^{3/2}g_m^{5/2}n_0^{3/2}\left[\dfrac{1}{16\sqrt{2}}\int_0^{2\pi}d\phi_k\int_0^\pi \sin\theta_k d\theta_k\left\{2[\beta+\mathcal{F}(\theta_k,\phi_k,\alpha)] k_c(\theta_k,\phi_k)^2\{5k_c(\theta_k,\phi_k)\right.\right.\nonumber\\
	&&-2\sqrt{2[\beta+\mathcal{F}(\theta_k,\phi_k,\alpha)]+k_c(\theta_k,\phi_k)^2}\}+6k_c(\theta_k,\phi_k)^4\{k_c(\theta_k,\phi_k)-\sqrt{2(\beta+\mathcal{F}(\theta_k,\phi_k,\alpha))+k_c(\theta_k,\phi_k)^2}\}\nonumber\\
	&&\left.\left.\hspace{2cm}+[\beta+\mathcal{F}(\theta_k,\phi_k,\alpha)]^2\{16\sqrt{2[\beta+\mathcal{F}(\theta_k,\phi_k,\alpha)]+k_c(\theta_k,\phi_k)^2}-15k_c(\theta_k,\phi_k)\}\right\}\right]
\end{eqnarray}

\section{Calculation of Lagrangian and Equations of motion}
Introducing the time dependent variational Gaussian ansatz,
\begin{eqnarray}
	\psi({\bm r}, t)&=&\dfrac{1}{\pi^{3/4}\sqrt{L_xL_yL_z}}\exp\left[-\dfrac{x'^2}{2L_x^2}-\dfrac{y^2}{2L_y^2}-\dfrac{z'^2}{2L_z^2}\right.\nonumber\\
	&&\hspace{1cm}\left.+ix'^2\beta_{x}+iy^2\beta_y+iz'^2\beta_{z}+ix'z'\beta_{xz}\right],
	\label{vg}
\end{eqnarray}
where $x'=x\cos\theta-z\sin\theta$, $z'=x\sin\theta+z\cos\theta$  into the Lagrangian density describing a DDBEC, 

	\begin{eqnarray}
		\mathcal{L}= \dfrac{i\hbar}{2}\left(\psi\dot\psi^*-\dot\psi\psi^*\right)+\dfrac{\hbar^2}{2m}|\nabla\psi|^2+\dfrac{g}{2}|\psi|^4+\dfrac{1}{2}|\psi|^2\!\! \int \!d^3r'V_d(r-r')|\psi(r')|^2+
		\dfrac{2M^{3/2}}{15\pi^3\hbar^3N}
		g_m^{5/2}|\psi|^5\!\! \int \!\! d\Omega_k[\beta+\mathcal{F}(\theta_k,\phi_k,\alpha)]^{5/2}
	\end{eqnarray}
we obtain the Lagrangian $L=\int d^3r \mathcal{L}$:
\begin{eqnarray}
	L&=& \dfrac{\hbar}{2}(L_x^2\dot{\beta_x}+L_y^2\dot{\beta_y}+L_z^2\dot{\beta_z})+\dfrac{\hbar}{2}(L_x^2-L_z^2)\beta_{xz}\dot{\theta}+\dfrac{\hbar^2}{4M}\left(\dfrac{1}{L_x^2}+\dfrac{1}{L_y^2}+\dfrac{1}{L_z^2}+(4\beta_x^2+\beta_{xz}^2)L_x^2+4\beta_y^2L_y^2+(4\beta_z^2+\beta_{xz}^2)L_z^2\right)\nonumber\\
	&&\hspace{1cm} +\dfrac{g}{2(2\pi)^{3/2}L_x L_y L_z}+\dfrac{g_m}{2}\int \dfrac{d^3k}{(2\pi)^3}\mathcal{F}(\theta_k,\phi_k,\alpha)n^2(k)+\left(\dfrac{2}{5}\right)^{5/2}\dfrac{(m/\hbar)^{3/2}}{3N\pi^{21/4}}\dfrac{g_m^{5/2}\int d\Omega_k[\beta+\mathcal{F}(\theta_k,\phi_k,\alpha)]^{5/2}}{(L_xL_yL_z)^{3/2}}
\end{eqnarray}
where $n(k)$ is the Fourier transform of the density $n(r)=|\psi(r)|^2$. The Euler-Lagrange equations for $\beta$ variables are: $\beta_{xz}=\dfrac{M}{\hbar}\left(\dfrac{L_z^2-L_x^2}{L_z^2+L_x^2}\right)$ and $\beta_i=\dfrac{M}{2\hbar}\dfrac{\dot{L}_i}{L_i}$ with $i\in \{x, y, z\}$. After the transformation $L\to L-\dfrac{d}{dt}\left(\dfrac{\hbar}{2}\sum_i\beta_iL_i^2 \right)$ we get,

	\begin{eqnarray}
		L&=& \dfrac{\hbar^2}{4M}\left(\dfrac{1}{L_x^2}+\dfrac{1}{L_y^2}+\dfrac{1}{L_z^2}\right)-\dfrac{M}{4}(\dot{L_x}^2+\dot{L_y}^2+\dot{L_z}^2)-\dfrac{M}{4}\dfrac{(L_x^2-L_z^2)^2}{(L_x^2+L_z^2)}\dot{\theta}^2\nonumber\\
		&& +\dfrac{g}{2(2\pi)^{3/2}L_x L_y L_z}+\dfrac{g_m}{2}\int \dfrac{d^3k}{(2\pi)^3}\mathcal{F}(\theta_k,\phi_k,\alpha)n^2(k)+\left(\dfrac{2}{5}\right)^{5/2}\dfrac{(M/\hbar)^{3/2}}{3N\pi^{21/4}}\dfrac{g_m^{5/2}\int d\Omega_k[\beta+\mathcal{F}(\theta_k,\phi_k,,\alpha)]^{5/2}}{(L_xL_yL_z)^{3/2}}
	\end{eqnarray}
The equations of motion, $\dfrac{d}{dt}\left(\dfrac{\partial L}{\partial \dot{Q}}\right)-\dfrac{\partial L}{\partial Q}=0$ with $Q\in \{L_x, L_y, L_z, \theta\}$ are

\begin{eqnarray}
	M\ddot{L}_x&=&\dfrac{\hbar^2}{ML_x^3}+\dfrac{ML_x\dot{\theta}^2}{(L_x^2+L_z^2)^2}(L_x^4+2L_x^2L_z^2-3L_z^4)+\dfrac{g}{(2\pi)^{3/2}L_x^2L_yL_z}-g_m\dfrac{\partial}{\partial L_x}\int \dfrac{d^3k}{(2\pi)^3}\mathcal{F}(\theta_k,\phi_k,\alpha)n^2(k)\nonumber\\
	&&\hspace{2cm}+\left(\dfrac{2}{5}\right)^{5/2}\dfrac{(M/\hbar)^{3/2}}{N\pi^{21/4}}\dfrac{g_m^{5/2}\int d\Omega_k[\beta+\mathcal{F}(\theta_k,\phi_k,\alpha)]^{5/2}}{L_x(L_xL_yL_z)^{3/2}}\label{el1}\\
	M\ddot{L}_y&=&\dfrac{\hbar^2}{ML_y^3}+\dfrac{g}{(2\pi)^{3/2}L_xL_y^2L_z}-g_m\dfrac{\partial}{\partial L_y}\int \dfrac{d^3k}{(2\pi)^3}\mathcal{F}(\theta_k,\phi_k,\alpha)n^2(k)+\left(\dfrac{2g_m}{5}\right)^{5/2}\dfrac{(M/\hbar)^{3/2}}{N\pi^{21/4}}\dfrac{\int d\Omega_k[\beta+\mathcal{F}(\theta_k,\phi_k,\alpha)]^{5/2}}{L_y(L_xL_yL_z)^{3/2}}\\
	M\ddot{L}_z&=&\dfrac{\hbar^2}{ML_z^3}+\dfrac{ML_z\dot{\theta}^2}{(L_x^2+L_z^2)^2}(L_z^4+2L_x^2L_z^2-3L_x^4)+\dfrac{g}{(2\pi)^{3/2}L_xL_yL_z^2}-g_m\dfrac{\partial}{\partial L_z}\int \dfrac{d^3k}{(2\pi)^3}\mathcal{F}(\theta_k,\phi_k,\alpha)n^2(k)\nonumber\\
	&&\hspace{2cm}+\left(\dfrac{2}{5}\right)^{5/2}\dfrac{(M/\hbar)^{3/2}}{N\pi^{21/4}}\dfrac{g_m^{5/2}\int d\Omega_k[\beta+\mathcal{F}(\theta_k,\phi_k,\alpha)]^{5/2}}{L_z(L_xL_yL_z)^{3/2}}\\
	\dfrac{(L_x^2-L_z^2)^2}{L_x^2+L_z^2}M\ddot{\theta}&=&-\dfrac{\partial}{\partial \theta}\left(g_m\int \dfrac{d^3k}{(2\pi)^3}\mathcal{F}(\theta_k,\phi_k,\alpha)n^2(k)\right)-2M\dot{\theta}\dfrac{(L_x^2-L_z^2)}{(L_x^2+L_z^2)^2}[L_x\dot{L_x}(3L_z^2+L_x^2)-L_z\dot{L_z}(3L_x^2+L_z^2)].
\end{eqnarray}

At the equilibrium the first derivatives vanishes. Thus, for the vicinity of the equilibrium point we can approximately write the equations of motion as $d^2Q'/dt^2=-\partial V_{eff}/\partial Q'$ with $Q'\in\{x', y, z', \theta'\}$ and $\theta'=\left([(L_x^0)^2-(L_z^0)^2]/\sqrt{(L_x^0)^2+(L_z^0)^2}\right)\theta$, 
where,

	\begin{eqnarray}
		V_{eff}=\frac{\hbar^2}{2M}\sum_i\frac{1}{L_i^2}+\frac{g}{(2\pi)^{3/2}L_xL_yL_z}+g_m\int \dfrac{d^3k}{(2\pi)^3}\mathcal{F}(\theta_k,\phi_k,\alpha)n^2(k)+2\left(\dfrac{2}{5}\right)^{5/2}\dfrac{(M/\hbar)^{3/2}}{3N\pi^{21/4}}\dfrac{g_m^{5/2}\int d\Omega_k[\beta+\mathcal{F}(k,\alpha)]^{5/2}}{(L_xL_yL_z)^{3/2}}
	\end{eqnarray}

The equilibrium widths and the orientation of the droplet are obtained by minimizing the effective potential, $V_{eff}$. Further, by linearizing the equations of motion around the equilibrium values we obtain the low-lying excitation frequencies.

\section{$N$-dependence of the droplet widths: Thomas-Fermi-Gaussian ansatz}
\subsection{$\alpha=0$: Cigar (Q1D) droplet}
3D numerical calculations revealed that at $\alpha=0$, in the Q1D regime we have a cigar-shaped droplet with almost a Gaussian shape in the radial direction and a TF-like profile in the axial direction. Thus, choosing a hybrid ansatz of the form:
\begin{eqnarray}
	n(\rho, z, t)&=& \dfrac{3}{4\pi L_\rho^2 R_z}\left(1-\dfrac{z^2}{R_z^2}\right)\exp\left(-\dfrac{\rho^2}{L_\rho^2}\right)e^{-i\mu t},
\end{eqnarray}
where $R_z$ is the TF radius and $L_{\rho}=L_x=L_y$ is the Gaussian width in the $xy$-plane. Using it in the generalized NLGPE [see Eq. (8) in the main text], neglecting the kinetic energy along the $z$-direction and then integrating over the $xy$  plane, we get:

	\begin{eqnarray}
		\mu&=&\dfrac{\hbar^2}{2ML_\rho^2}+\dfrac{3g}{8\pi L_\rho^2R_z}\left(1-\dfrac{z^2}{R_z^2}\right)+\dfrac{\sqrt{3}g_m^{5/2}}{20 N\pi^{9/2}}\left(\dfrac{m}{R_z L_\rho^2\hbar^2}\right)^{3/2}\int d\Omega_k[\beta+\mathcal{F}(\theta_k,\phi_k,0)]^{5/2}\left(1-\dfrac{z^2}{R_z^2}\right)^{3/2}\nonumber\\
		&&\hspace{0cm}+\left(\dfrac{\pi g_m}{R_z^3}\right)\int\dfrac{dk_z}{2\pi}\int\dfrac{dk_x}{2\pi}\int\dfrac{dk_y}{2\pi}\left(\dfrac{3k_z^2}{k_x^2+k_y^2+k_z^2}-1\right)e^{-k_\rho^2L_\rho^2/2}\dfrac{4(\sin k_zR_z-k_zR_z\cos k_zR_z)}{k_z^3}e^{iz k_z}.
	\end{eqnarray}

Upon expanding around $z=0$ and equating the coefficients of the $z^0$ and $z^2$ terms we get respectively,

	\begin{eqnarray}
		\mu&=&\dfrac{\hbar^2}{2ML_\rho^2}+\dfrac{3}{8\pi L_\rho^2R_z}\left(g-\dfrac{4\pi g_m}{3}\right)+\dfrac{\sqrt{3}g_m^{5/2}}{20 N\pi^{9/2}}\left(\dfrac{M}{R_z L_\rho^2\hbar^2}\right)^{3/2}\int d\Omega_k[\beta+\mathcal{F}(\theta_k,\phi_k,0)]^{5/2}\nonumber\\
		&&-\dfrac{6g_m}{4R_z^2L_\rho}\left(\sqrt{2\pi}e^{R_z^2/2L_\rho^2}\text{erfc}\left[\dfrac{R_z}{\sqrt{2}L_\rho}\right]-\dfrac{\pi L_\rho}{R_z}\text{erfi}\left[\dfrac{R_z}{\sqrt{2}L_\rho}\right]+\dfrac{R_z}{L_\rho}~_pF_q\left(\{1,1\};\left\{\frac{3}{2},2\right\};\frac{R_z^2}{2L_\rho^2}\right)\right)
	\end{eqnarray}

and

	\begin{eqnarray}
		&&\dfrac{1}{2\pi L_\rho^2R_z}\left(\beta-\dfrac{4\pi}{3}\right)+\dfrac{\sqrt{3N}}{10\pi^3L_\rho^3}\left(\dfrac{3a_m}{\pi R_z}\right)^{3/2}\int d\Omega_k[\beta+\mathcal{F}(\theta_k,\phi_k,0)]^{5/2}\nonumber\\
		&&+\left(\dfrac{2}{R_zL_\rho^2}-\dfrac{\sqrt{2\pi}}{L_\rho^3}e^{\dfrac{R_z^2}{2L_\rho^2}}\text{erfc}\left[\dfrac{R_z}{\sqrt{2}L_\rho}\right]\right)-\dfrac{1}{\pi R_z^3}\int dx ~e^{\dfrac{x^2 L_\rho^2}{2R_z^2}}\Gamma\left[0,\dfrac{x^2L_\rho^2}{2R_z^2}\right]x^2\cos x=0
		\label{wth}
	\end{eqnarray}

where $\rm {erfi}()$ and $\rm {erfc}()$ are the imaginary and complimentary error functions respectively,  $~_pF_q\left(\right)$ is the generalized hypergeometric function and $\Gamma []$ is the incomplete gamma function. In Eq. (\ref{wth}) out of the four terms, the first term comes from the effective short-range interactions including those from the DDI, second is from the LHY correction and the last two terms emerge from the non-local character of the DDI which can be neglected in the limit $R_z\gg L_{\rho}$, and finally we get:
\begin{eqnarray}
	R_z\simeq\dfrac{81a_m^3 N}{25\pi^7L_{\rho}^2\left(4\pi/3-\beta\right)^2}\left[\int d\Omega_k\left[\beta+\mathcal{F}(\theta_k,\phi_k,0)\right]^{5/2}\right]^2.
\end{eqnarray}
Similarly, for $\alpha=\pi/2$, in the Q2D regime, using an ansatz 
\begin{eqnarray}
	n(\rho, z, t)&=&\dfrac{3}{4\sqrt{\pi}R_{\perp}^2 L_y}\left(1-\dfrac{\rho^2}{R_{\perp}^2}\right)e^{-y^2/L_y^2}e^{-i\mu t},
\end{eqnarray}
and following an identical procedure as we did for $\alpha=0$, finally we get
\begin{eqnarray}
	R_\perp\simeq\dfrac{9a_m^{3/2}}{2\pi^{13/4}}\sqrt{\dfrac{N}{5L_y}}\dfrac{\int d\Omega_k[\beta+\mathcal{F}(\theta_k,\phi_k,\pi/2)]^{5/2}}{\left(8\pi/3-\beta\right)}.
\end{eqnarray}

\end{document}